\documentclass[final]{siamltex}

\usepackage{rotating}
\usepackage{amsmath}
\usepackage{amsfonts} 
\usepackage{amssymb}

\def\R{{\mathbb{R}}}
\def\RD{{\mathbb{R}^2}}
\def\TD{{\mathbb{T}^2}}

\def\Div{\textup{div\,}}
\def\BV{\mathcal{BV}}
\def\S{\mathcal{S}}

\title{Multiscale texture separation}
\author{J\'er\^ome Gilles \thanks{University of California, Los Angeles (UCLA), Department of Mathematics, 520 Portola Plaza, Los Angeles, CA 90095 ({\tt jegilles@math.ucla.edu})}}

\begin{document}

\maketitle

\begin{abstract}
In this paper, we investigate theoretically the behavior of Meyer's image cartoon + texture decomposition model. Our main results is a new theorem 
which shows that, by combining the decomposition model and a well chosen Littlewood-Paley filter, it is possible to extract almost perfectly a certain class 
of textures. This theorem leads us to the construction of a parameterless multiscale texture separation algorithm. Finally, we propose to extend this 
algorithm into a directional multiscale texture separation algorithm by designing a directional Littlewood-Paley filter bank. Several experiments show 
the efficiency of the proposed method both on synthetic and real images.
\end{abstract}



\section{Introduction}
In the last decade, based on the work of Meyer \cite{Meyer2001}, many papers were published on cartoon + textures decomposition models for images. Some 
of them address numerical issues \cite{Aujol2005a,Aujol2003a,Luminita2003a}, others the modeling aspects \cite{Aujol2005b,Aujol2006b,Gilles2007,
Triet2005a,Osher2003a}, a few focus on applications \cite{Aujol2003a,Aujol2006a,Aujol2006c,Aujol2004,Aujol2004a,Gilles2010a}. Finally 
a very small number of publications tries to characterize the solutions of such models with respect to the choice of the parameters \cite{Aujol2006b,Gilles2010a}.
In \cite{Tadmor2004}, Tadmor et al. proposed a multiscale cartoon representation of an image. They iterate the Rudin-Osher-Fatemi model 
to consecutively extract objects that belong to different scales. This work does not deal with textures at different scales but it is interesting as it provides the relationship between the notion of scale and 
the choice of the regularizing parameter of the algorithm. \\
Initially, the idea of decomposing an image is to separate different kind of information: objects and textures. Indeed, in term of 
analysis (like, for example a segmentation task) it is useful to have separated specific information. For example, a classical way to analyze textures is 
to use some wavelet type filtering and then use the obtained coefficients to build a feature vector which can be provided to some classifier. While the idea of 
extracting such texture feature vectors directly from the texture part of the decomposition seems natural, no publication 
really addresses the construction of a well-defined texture separation algorithm based on decomposition models. In this paper, we investigate the 
possibility to optimally combine image decomposition and a well chosen filtering to extract specific textures in an image. Based on this result, we 
propose a multiscale texture separation algorithm.\\

To do this program, we propose a general formulation to decompose an image $f\in L^2(\RD)$ into three parts $u,v$ and $w$. The first one represents the 
objects contained in $f$, the second part is a residual term while the last one models the highest oscillating parts in the image.

In our work, we model $u,v$ and $w$ by three different functional spaces, 
$BV,L^2$ and the space $G$ which is, in a sense defined below, the dual of $BV$. Moreover, we use two complementary parameters $\lambda >0$ and $\mu>0$ to control the 
behavior of the algorithm. Finally, the optimal decomposition must minimize
\begin{equation}
J(u,v,w)=\|u\|_{BV}+\lambda\|v\|_{L^2}^2+\mu\|w\|_G
\end{equation}
over all possible decompositions $f=u+v+w$ of $f$.

We will see that this program works only if the parameters $\lambda>0$ and $\mu>0$ are, a posteriori, fixed accordingly to the processed image. 
If $\mu$ is too large, we necessarily have $w=0$ and the decomposition algorithm is equivalent to the well known Rudin-Osher-Fatemi (ROF) \cite{Rudin1992} algorithm. 
If $\mu$ is too small ($0<\mu<4\pi$) then we have $u=0$ and the algorithm degenerates.

The remainder of the paper is as follows. In section~\ref{sec:fsp} we recall the definition of the used function spaces and some of their properties. 
Section~\ref{sec:imdec} gives a detailed presentation of the decomposition model and recall some of its properties which will be useful to prove our main 
result. In section~\ref{sec3}, we prove a theorem which states that we can retrieve, almost perfectly, some specific textures from the texture part and 
a well chosen Littlewood-Paley filter. Section \ref{sec:noisy} provides a more precise result in the case of noisy images. A multiscale texture separation 
algorithm is proposed in section~\ref{sec:textsep} and is extended to a directional multiscale texture separation algorithm in section~\ref{sec:dmts}. 
We conclude this work in section~\ref{sec:conc}.

\section{The function spaces}\label{sec:fsp}
In this section, we recall the definition of the function spaces used in image decomposition models. In \cite{Meyer2001}, Meyer defined the decomposition 
idea on the basis of the ROF model \cite{Rudin1992} which uses the function space $BV$ (the space of functions of Bounded Variations) to model objects in an image where the norm on $BV$ is defined by 
$\|u\|_{BV}=\|u\|_{L^1}+|Du|$ where,

\begin{equation}\label{eq:deftv}
|Du|=\sup\left\{\int_{\Omega} u\Div\phi dx: \phi\in\mathcal{C}_c^{\infty}(\Omega,\R^N), |\phi|\leqslant 1\; \forall x\in\Omega \vphantom{\int}\right\}
\end{equation}

Meyer proposed 
to modify the ROF model by using dual concepts to characterize textures as oscillating patterns in an image.
But rigorously speaking, duals of $BV(\RD)$ or $L^{\infty}(\RD)$ do not exist. The problem vanishes if we keep the same norm and consider the closure of 
$\S(\RD)$ (Schwartz function class) in the studied space. For example, instead of $BV(\RD)$, we get a (closed) space $\BV\subset BV$ or instead of 
$L^{\infty}(\RD)$ we get $\mathcal{C}_0(\RD)$ (the space of continuous functions vanishing at infinity), etc.

The dual of $\BV$ is the space $G\subset\S'(\RD)$ (the dual space of $\S$). Meyer shows that functions or distributions $f\in G$ can be seen as the divergence of a vector field $F\in L^{\infty}(\RD)\times L^{\infty}(\RD)$. 
More precisely, the $G-$norm of $f\in G$, denoted $\|f\|_G$, is defined by
\begin{equation}\label{eq:gnorm1}
\|f\|_G=\inf \left\{\|F\|_{\infty};f=\Div F\right\}
\end{equation}
where
\begin{equation}
\|F\|_{\infty}=\left\| \left(|F_1(x)|^2+|F_2(x)|^2 \right)^{\frac{1}{2}}\right\|_{\infty}
\end{equation}
with $F=(F_1,F_2)$.

The question of $G$ dual has no meaning. In its strict sense, $G$ is not the dual of $BV$, and $BV$ is not the dual of $G$.

We will denote $G_0$ the closure of $L^2$ in $G$. A consequence of $BV\subset L^2$ is that $L^2\subset G$. Then the dual of $G_0$ is in $BV$.

In the following, we need to equip $G_0$ of the fanciful norm:

\begin{equation}\label{eq:normmu}
\|f\|_{\mu}=\inf \left\{\|u\|_{BV}+\mu\|v\|_G  \right\}
\end{equation}
where the decomposition is considered over all $f=u+v$ decompositions.

The dual space of $G_0$ (endowed with $\|f\|_{\mu}$) is $BV$ associated with the norm
\begin{equation}\label{eq:normbvmu}
\|f\|_{BV_{\mu}}=\sup \left\{\frac{\|f\|_{BV}}{\mu},\|f\|_G \right\};
\end{equation}
we are interested in the case wherein $\mu$ is larger than one. If $0<\mu\leqslant 4\pi$, the norm $\|f\|_{BV_{\mu}}$ simply is 
$\frac{1}{\mu}\|f\|_{BV}$ (this is a result of the isoperimetric inequality) while $\|f\|_{\mu}=\mu\|v\|_G$.

\section{The image decomposition algorithm}\label{sec:imdec}
We propose to decompose an image $f\in L^2(\RD)$ into three components $u,v$ and $w$. The first one represents the 
objects contained in $f$, the second part is a residual term while the last one models the highest oscillating parts in the image.

The variational algorithm providing the $f=u+v+w$ decomposition aims to minimize 
\begin{equation}\label{eq:modelinit}
\inf\left\{\|u\|_{BV}+\lambda\|v\|_{L^2}^2+\mu\|w\|_G\right\}.
\end{equation}

Naturally, we can rewrite this algorithm into an $f=g+h$ algorithm where we need to minimize
\begin{equation}
\inf \left\{\|g\|_E+\lambda\|h\|_{L^2}^2\right\}
\end{equation}
where the functional space $E$ is $G_0$ endowed with the norm given by Eq.~(\ref{eq:normmu}). We set $g=u+w$ and $h=v$ and then 
results for ``generalized'' ROF algorithms apply:

\begin{theorem}\label{theo1}
Let $V\subset L^2(\RD)$ a normed vector space. We suppose the norm $\|.\|_V$ has the following upper semi-continuity property: if $f_j\rightharpoonup f$ 
in $L^2$, then $\|f\|_V\leqslant\liminf \|f_j\|_V$.

Then we consider the unique optimal decomposition of $f\in L^2(\RD)$ into $u+v$ minimizing $\|u\|_V+\lambda\|v\|_2^2$. If $\|.\|_*$ denotes the dual norm 
of $V$, we have
\begin{remunerate}
\item if $\|f\|_*\leqslant\frac{1}{2\lambda}$, then $u=0$ and $f=v$,
\item if $\|f\|_*>\frac{1}{2\lambda}$, the optimal decomposition is characterized by
\begin{equation}
\|v\|_*=\frac{1}{2\lambda}, \qquad \int uvdx=\|u\|_V\|v\|_*.
\end{equation}
\end{remunerate}
\end{theorem}

The proof of this theorem is based on an analysis of the optimal pair solution of the ROF model, see Lemma 3,4 and theorem 3 in \cite{Meyer2001} (a more general form of this theorem can be found in \cite{aubin1984applied}). 
Theorem~\ref{theo1} and Eq.~(\ref{eq:normbvmu}) yield into the important following result (originally proven in \cite{Gilles2010a}) which characterizes the 
solutions of the decomposition model with respect to the input parameters.

\begin{theorem}\label{theo2}
If $\|f\|_{BV}\leqslant\frac{\mu}{2\lambda}$ and if $\|f\|_G\leqslant\frac{1}{2\lambda}$, then the optimal decomposition of $f$ is given by $u=0,v=f$ 
and $w=0$.

If either $\|f\|_{BV}>\frac{\mu}{2\lambda}$ or $\|f\|_{G}>\frac{1}{2\lambda}$, then necessarily we have
\begin{equation}
\|v\|_{BV}=\frac{\mu}{2\lambda}\qquad \text{and}\qquad \|v\|_G\leqslant\frac{1}{2\lambda}
\end{equation}
or
\begin{equation}
\|v\|_{BV}\leqslant\frac{\mu}{2\lambda}\qquad \text{and}\qquad \|v\|_G=\frac{1}{2\lambda}
\end{equation}
and $\langle u+w,v\rangle=\frac{1}{2\lambda}\|u+w\|_{\mu}$.
\end{theorem}

This theorem shows us that if $\lambda>0$ is too small and if $f\in BV$, we are in the first case and the optimal decomposition is obvious: the whole image is captured by $v$
(the term $\lambda\|v\|_2^2$ is not enough penalized). If $f$ is enough oscillating to have 
$\|f\|_G\leqslant\frac{1}{2\lambda}$ and if $\mu$ is large, we also are in the case where $f$ is entirely captured by $v$: the term $\mu\|w\|_G$ is too much 
penalizing.

The following Theorem~\ref{theo3} is a variant version of Theorem~\ref{theo2}. It is not a corollary of Theorem~\ref{theo2}.

\begin{theorem}\label{theo3}
Assume $f\in BV$ and $\|f\|_{BV}\leqslant\frac{\mu}{4\lambda}$. Then the optimal decomposition $f=u+v+w$ necessarily verifies $w=0$. 
There are no textures.
\end{theorem}

In other terms, if $\mu$ is too large, then $\mu\|w\|_G$ is too penalizing.

\begin{proof}
We have $\|u\|_{BV}\leqslant\|f\|_{BV}$ (compare $f=u+v+w$ to $f=f+0+0$). Consequently, we have 
$\|v+w\|_{BV}=\|f-u\|_{BV}\leqslant 2\|f\|_{BV}\leqslant\frac{\mu}{2\lambda}$. Consider $u$ is fixed and set $\sigma=v+w$.  Then for all  
fixed $u$, we are supposed to minimize $\lambda\|v\|_2^2+\mu\|w\|_G$. We apply the general theory. The dual norm of the $G-$norm is the $BV-$norm. 
Then if $\|\sigma\|_{BV}\leqslant\frac{\mu}{2\lambda}$, the optimal decomposition of $\sigma$ is given by $v=\sigma$ and $w=0$.
\qquad\end{proof}

\begin{corollary}
Under assumptions of Theorem~\ref{theo3}, the optimal decomposition of $f$ is provided by the ROF algorithm.
\end{corollary}

\section{Optimal texture separation}\label{sec3}

In this section, we study the behavior of the decomposition algorithm with respect to the presence of different oscillating patterns.
We consider the following image $f(x)=a(x)+b(x)\cos(\omega_1x+\varphi_1)+c(x)\cos(\omega_2x+\varphi_2)$ with $|\omega_1|\ll|\omega_2|$.

Firstly, note that if $\frac{\mu}{\lambda}\gg|\omega_2|$, then theorem \ref{theo3} applies and we get $w=0$ (because $\|f\|_{BV}\approx|\omega_2|$).

Secondly, let us examine the case where $1\leqslant\lambda\ll |\omega_1|\ll\frac{\mu}{\lambda}\ll |\omega_2|$. 
We will prove that, for some specific $\lambda$ and $\mu$, $w(x)$ is essentially equal to $c(x)\cos(\omega_2x+\varphi_2)$. We assume that $a,b,c$ 
are $\mathcal{C}^1$ functions with compact support.

Let us begin by evaluating the energy of the corresponding decomposition. It is given by $J_0(f)=\|a\|_{BV}+\lambda\|b\|_2^2+\frac{\mu}{|\omega_2|}$ 
and is bounded by $C\lambda$.

Thus we have $\|u\|_{BV}\leqslant C\lambda$, $\lambda\|v\|_2^2\leqslant C\lambda$. We apply theorem \ref{theo2} and we get 
$\|f\|_{BV}\simeq C|\omega_2|$ which is much larger than $\frac{\mu}{\lambda}$. We necessarily have $\|v\|_{BV}\leqslant\frac{\mu}{2\lambda}$. Finally, we
get
\begin{equation}\label{eq:fwbound}
\|f-w\|_{BV}\leqslant C\frac{\mu}{\lambda}.
\end{equation}

In order to prove the following main theorem, we start to prove the next two useful lemmas (the hat symbol stands for the Fourier transform).

\begin{figure}[t!]
\includegraphics[width=\textwidth,height=4cm]{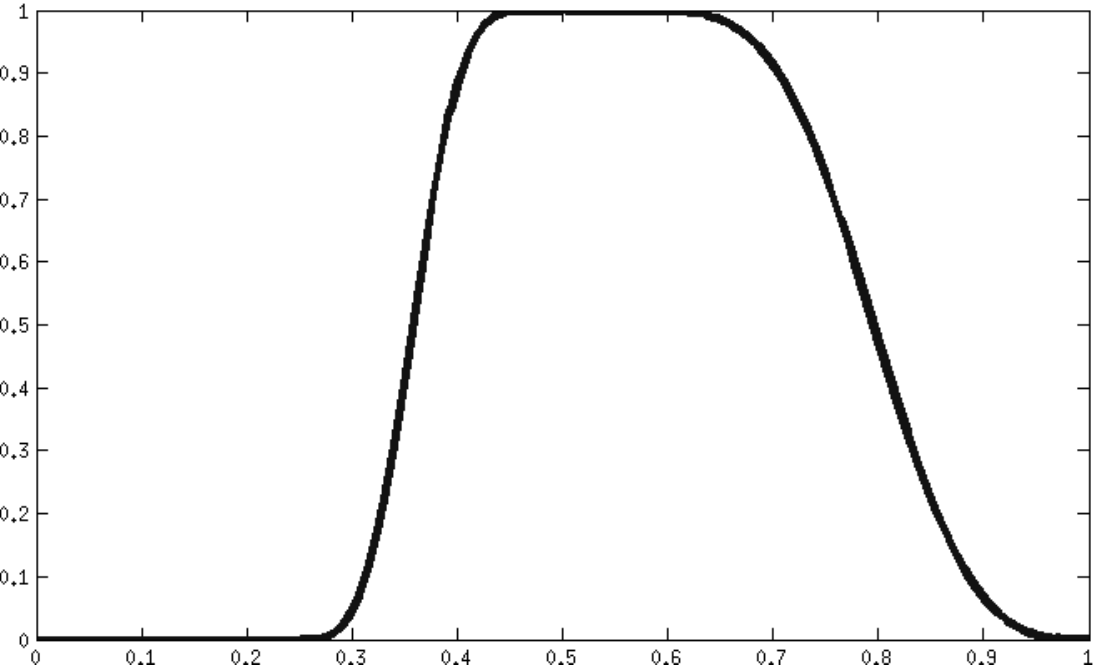}
\caption{Illustration of the Fourier transform magnitude of a $\Delta_j$ operator in one dimension.}
\label{fig:littlewood}
\end{figure}

\begin{definition}\label{def:LP}
The Littlewood-Paley filter associated with scale $j$, denoted $\Delta_j$, is defined by
\begin{align}
\hat{\Delta}_j(\xi)=
\begin{cases}
1 \qquad \text{if}\quad 2^{j-1}\leqslant\xi\leqslant 2^{j} \\
0 \qquad \text{if}\quad \xi\leqslant 2^{j-2} \qquad \text{or} \qquad 2^{j+1}\leqslant \xi.
\end{cases}
\end{align}
\end{definition}
Figure~\ref{fig:littlewood} sketches the magnitude of the Fourier transform of such operator.

\begin{lemma}\label{lem:delta}
Let $E$ a functional space, $\|.\|_E$ its associated norm which we will assume is translation invariant (like $BV, L^p,\ldots$), $\Delta_j$ 
is a Littlewood-Paley operator associated to the function $\psi$. Then for all 
function $f\in E$
\begin{equation}
\|\Delta_j[f]\|_E\leqslant C\|f\|_E \qquad \text{where}\quad C=\|\psi\|_{L^1}.
\end{equation}
\end{lemma}
\begin{proof}
The proof is straightforward:
\begin{align}
\|\Delta_j[f]\|_E&\leqslant \left\| \int 2^{2j}\psi(2^jy)f(x-y)dy\right\|_E \\
&\leqslant\int 2^{2j}|\psi(2^jy)|dy \|f\|_E\\
&\leqslant\|\psi\|_{L^1}\|f\|_E.
\end{align}
\qquad\end{proof}


\begin{lemma}\label{lem:bvl1}
If $\hat{f}$ is supported by $R\leqslant|\xi|\leqslant 3R$ with $R\gg1$ then $\|f\|_{BV}\approx R\|f\|_{L^1}$.
\end{lemma}
This lemma is a direct consequence of Bernstein's inequalities.\\

Now let us prove the following theorem which asserts that a particular texture can be separated from the rest of the image.
\begin{theorem}\label{th:sep}
If $f(x)=a(x)+b(x)\cos(\omega_1x+\varphi_1)+c(x)\cos(\omega_2x+\varphi_2)$ and if we assume that 
$1\leqslant\lambda\ll |\omega_1|\ll\frac{\mu}{\lambda}\ll |\omega_2|$ then there exists a constant $C$ such that $f=u+v+w$ verifies, 
for a certain integer $j$,
\begin{equation}
\|\Delta_j[w](x)-c(x)\cos(\omega_2x+\varphi_2)\|_{L^1}\leqslant C\frac{\mu}{\lambda|\omega_2|}.
\end{equation}
\end{theorem}
\begin{proof}
We denote $r=u+v$, $g(x)=a(x)+b(x)\cos(\omega_1x+\varphi_1)$, $W(x)=c(x)\cos(\omega_2x+\varphi_2)$, then $f=g+W=r+w$. 
It is easy to see that $\|g\|_{BV}\leqslant C_1|\omega_1|\leqslant C\frac{\mu}{\lambda}$. Moreover,
\begin{align}
\|w-W\|_{BV}&\leqslant\|f-r-(f-g)\|_{BV}\\
&\leqslant \|r\|_{BV}+\|g\|_{BV}.
\end{align}
But, because of Eq.~(\ref{eq:fwbound}), we have $\|r\|_{BV}\leqslant C_2\frac{\mu}{\lambda}$, which finally conducts to 
\begin{equation}\label{eq:wW}
\|w-W\|_{BV}\leqslant C\frac{\mu}{\lambda}.
\end{equation}
Now, let $\Delta_j$ be a Littlewood-Paley filtering operator defined as in Definition.~\ref{def:LP}.
We assume that the scale $j$ is chosen such that $2^{j-1}\leqslant\omega_2\leqslant 2^{j}$.
We denote $\Delta_j(w-W)=w_j-W_j$. Then equation Eq.~(\ref{eq:wW}) and lemma~\ref{lem:delta} yields
\begin{equation}
\|w_j-W_j\|_{BV}\leqslant C\frac{\mu}{\lambda}.
\end{equation}
Note that $\widehat{(w_j-W_j)} $ is supported by $2^{j-2}\leqslant\xi\leqslant 2^{j+1}$. We apply lemma~\ref{lem:bvl1} which provides
\begin{equation}\label{eq:wi1}
\|w_j-W_j\|_{L^1}\leqslant C\frac{\mu}{\lambda|\omega_2|}.
\end{equation}
Otherwise we have
\begin{equation}
\|W_j-W\|_{L^1}\leqslant C\frac{\mu}{\lambda|\omega_2|}.
\end{equation}
Indeed, if we decompose $c(x)$ by a paraproduct algorithm: $c(x)=c_j(x)+\gamma_j(x)$ where
\begin{equation}
\hat{c}_j(\xi)=\hat{c}(\xi)\hat{\varphi}\left(\frac{\xi}{2^j}\right)
\end{equation}
and where $\varphi$ is a lowpass filter: $\hat{\varphi}(\xi)\neq 0$ if $|\xi|\leqslant\frac{1}{N}$ for a fixed $N\gg 1$. Then 
$c(x)\cos(\omega_2 x+\varphi_2)=c_j(x)\cos(\omega_2 x+\varphi_2)+\gamma_j\cos(\omega_2 x+\varphi_2)$. 
Taking the inverse Fourier transform, we have ($n$ being the dimension)
\begin{equation}
c_j=\left(\frac{1}{2\pi}\right)^n\int e^{\imath\xi x}\hat{c}(\xi)\hat{\varphi}\left(\frac{\xi}{2^j}\right) d\xi
\end{equation}
then
\begin{equation}
c_je^{\imath(\omega_2w+\varphi_2)}=\left(\frac{1}{2\pi}\right)^n\int e^{\imath(\xi+\omega_2) x+\varphi_2}\hat{c}(\xi)\hat{\varphi}\left(\frac{\xi}{2^j}\right) d\xi.
\end{equation}
From the initial assumption, we know that $2^{j-1}\leqslant |\omega_2|\leqslant 2^j$ and from the definition of $\hat{\varphi}$ given above, we know that
$\hat{\varphi}(\frac{\xi}{2^j})\neq 0$ if $\left|\frac{\xi}{2^j}\right|\leqslant\frac{1}{N}$. Finally we get that for
\begin{equation}
\left(1-\frac{1}{N}\right)2^{j-1}\leqslant|\xi+\omega_2|\leqslant\left(1+\frac{1}{N}\right)2^j,
\end{equation}
$\Delta_j$ is the identity ($\Delta_j=I$). As $N$ is considered much larger than one, everything happen in the dyadic ring between 
$2^{j-1}$ and $2^j$. Notably, this means that $\Delta_j[c_j\cos(\omega_2x+\varphi_2)]=c_j(x)\cos(\omega_2x+\varphi_2)$.
Then we have 
\begin{align}
\Delta_j[W]-W&=\Delta_j[c_j(x)\cos(\omega_2x+\varphi_2)]+\Delta_j[\gamma_j(x)\cos(\omega_2x+\varphi_2)]\\\notag
&-c_j(x)\cos(\omega_2x+\varphi_2)-\gamma_j(x)\cos(\omega_2x+\varphi_2)\\
&=\Delta_j[\gamma_j(x)\cos(\omega_2x+\varphi_2)]-\gamma_j(x)\cos(\omega_2x+\varphi_2)
\end{align}
Now we take the $L^1-$norm:
\begin{align}
\|\Delta_j[W]-W\|_{L^1}&=\|\Delta_j[\gamma_j(x)\cos(\omega_2x+\varphi_2)]-\gamma_j(x)\cos(\omega_2x+\varphi_2)\|_{L^1} \\
&\leqslant\|\Delta_j[\gamma_j(x)\cos(\omega_2x+\varphi_2)]\|_{L^1}+\|\gamma_j(x)\cos(\omega_2x+\varphi_2)\|_{L^1} \\
\end{align}
Lemma~\ref{lem:delta} gives $\|\Delta_j[\gamma_j(x)\cos(\omega_2x+\varphi_2)]\|_{L^1}\leqslant C_3\|\gamma_j(x)\cos(\omega_2x+\varphi_2)\|_{L^1}$ and 
consequently
\begin{align}
\|\Delta_j[W]-W\|_{L^1}&\leqslant C_3\|\gamma_j(x)\cos(\omega_2x+\varphi_2)\|_{L^1}+\|\gamma_j(x)\cos(\omega_2x+\varphi_2)\|_{L^1} \\
&\leqslant C_4\|\gamma_j(x)\cos(\omega_2x+\varphi_2)\|_{L^1}\\
&\leqslant C_4\|\gamma_j\|_{L^1}.
\end{align}
But
\begin{align}
\gamma_j(x)=& c(x)-\int c(x-2^{-j}y)\varphi(y)dy\\
=&\int(c(x)-c(x-2^{-j}y))\varphi(y)dy.
\end{align}
If we denote $d_j(x,y)=c(x)-c(x-2^{-j}y)$ then the triangular inequality provides
\begin{equation}
\|\gamma\|_{L^1}\leqslant\int\|d_j(.,y)\|_{L^1}|\varphi(y)|dy,
\end{equation}
$\|d_j(.,y)\|_{L^1}$ meaning that the $L^1-$norm is taken with respect to the first variable. Let us assume that $c\in BV$, one property of $BV$ 
is that there exists a constant $C$ such that $\|c(x)-c(x+y)\|_{L^1}\leqslant C|y|$. We deduce that $\|d_j(.,y)\|_{L^1}\leqslant C2^{-j}|y|$ and finally 
$\|\gamma_j\|_{L^1}\leqslant C'2^{-j}$. We know that $j$ is chosen such that $|\omega_2|\leqslant 2^j$ and consequently 
$\|\gamma_j\|_{L^1}\leqslant\frac{C'}{|\omega_2|}$. This permits us to conclude that there exists a constant $C_5$ such that
\begin{equation}\label{eq:wi2}
\|W_j-W\|_{L^1}\leqslant\frac{C_5}{|\omega_2|}
\end{equation}
The combination of Eq.~(\ref{eq:wi1}) Eq.~(\ref{eq:wi2}) allows us to conclude that there exists a constant $C$ such that
\begin{equation}
\|w_j-W\|_{L^1}\leqslant C\frac{\mu}{\lambda|\omega_2|}
\end{equation}
This ends the proof.
\qquad\end{proof}

\begin{figure}[!h]
\begin{tabular}{cc}
\includegraphics[width=0.48\textwidth]{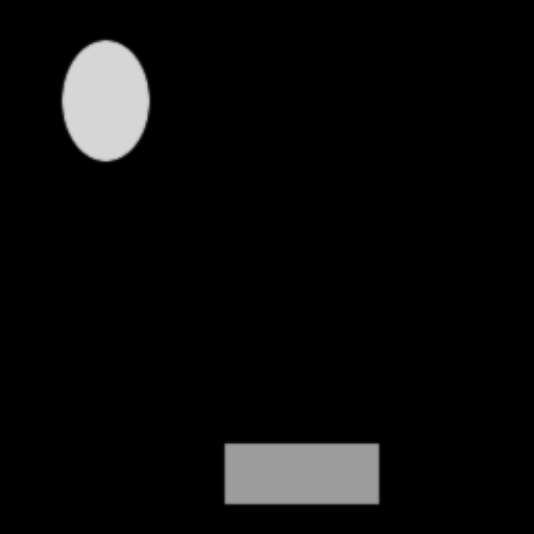} &
\includegraphics[width=0.48\textwidth]{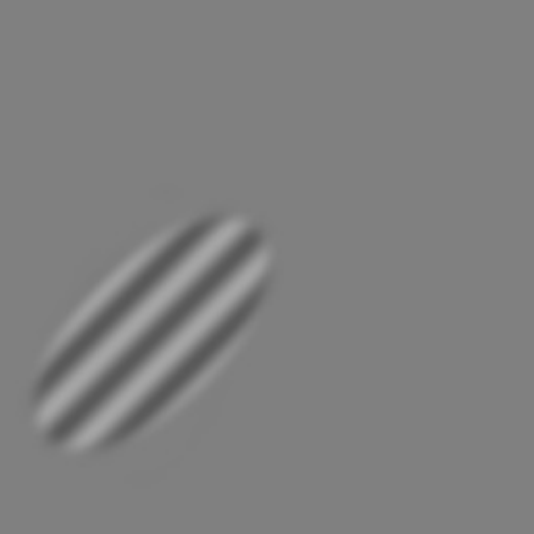} \\
$a(x)$ & $b(x)\cos(\omega_1x)$
\end{tabular}
\centering\includegraphics[width=0.48\textwidth]{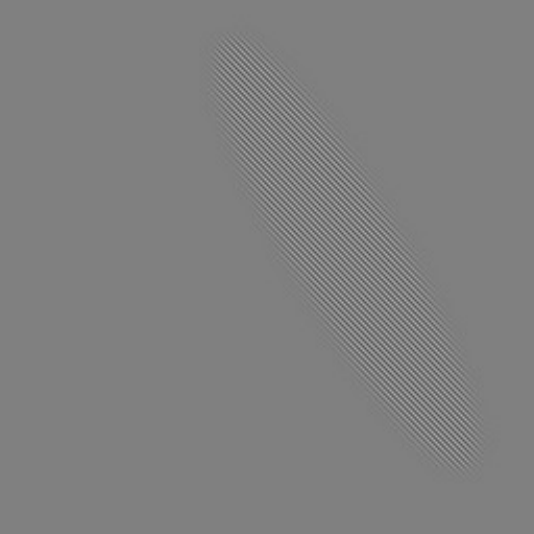} \\
$c(x)\cos(\omega_2x)$
\caption{Reference synthetic components.}
\label{fig:imref}
\end{figure}

\begin{figure}[!h]
\centering\includegraphics[width=0.48\textwidth]{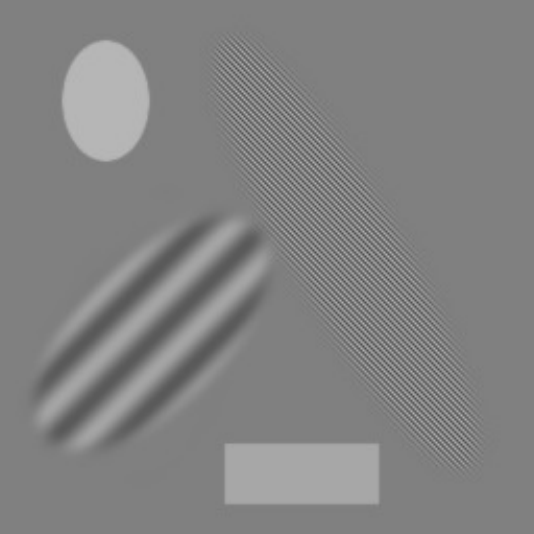}
\caption{The whole synthetic test image $f$.}
\label{fig:im}
\end{figure}

In order to verify this theorem experimentally, we build a synthetic image which contains the different components expected by the theorem assumptions.
The Little\-wood-Paley filter is implemented by using Meyer's wavelet.  
The test image is composed of a $BV$ type part and two different frequential components ($\omega_1=025.6 Rad/s$, $\omega_2=256 Rad/s$ and without any loss 
of generality $\varphi_1=\varphi_2=0$) defined over finite domains, see Fig.~\ref{fig:imref} and \ref{fig:im}. 
Accordingly to the theorem, we fix the parameters as follows: $\lambda=1$ and $\mu=100$. 

\begin{figure}[!h]
\begin{tabular}{cc}
\includegraphics[width=0.48\textwidth]{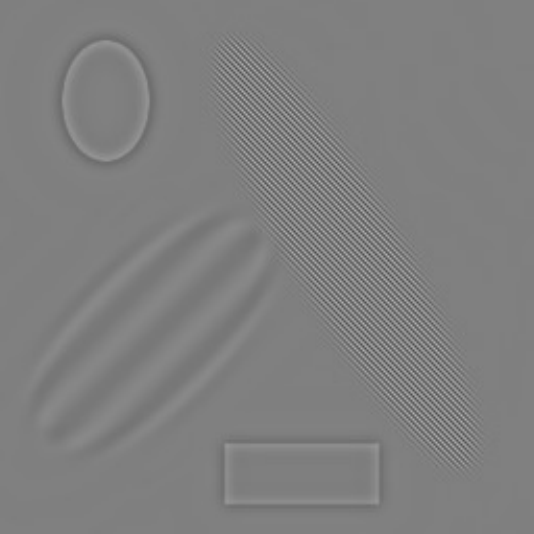} &
\includegraphics[width=0.48\textwidth]{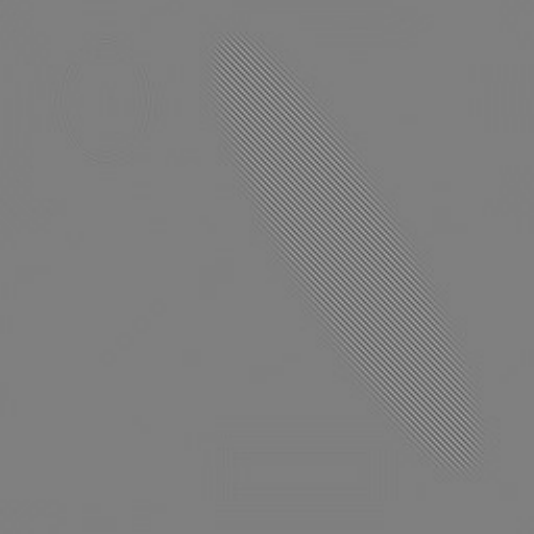}
\end{tabular}
\caption{Component $w$ provided by the decomposition and its Littlewood-Paley filtered version $\Delta_j[w](x)$.}
\label{fig:w}
\end{figure}

\begin{figure}[!h]
\begin{tabular}{cc}
\fbox{\includegraphics[width=0.45\textwidth]{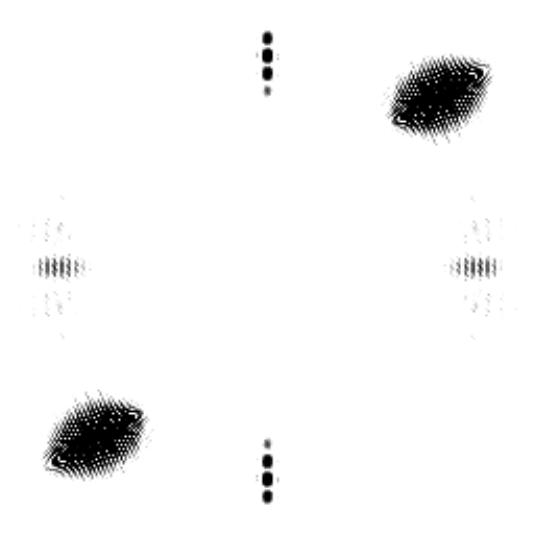}} &
\fbox{\includegraphics[width=0.45\textwidth]{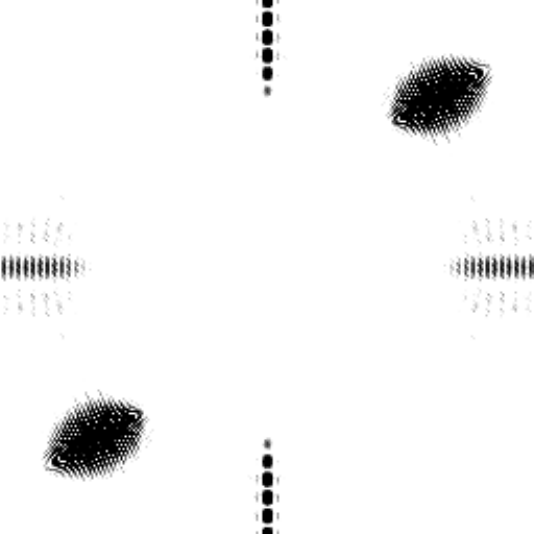}}
\end{tabular}
\caption{Log-amplitude of Fourier transforms of $\Delta_j[w](x)$ (left) and $\Delta_j[f](x)$ (right).}
\label{fig:ffts}
\end{figure}

\begin{figure}[!h]
\centering\includegraphics[width=0.48\textwidth]{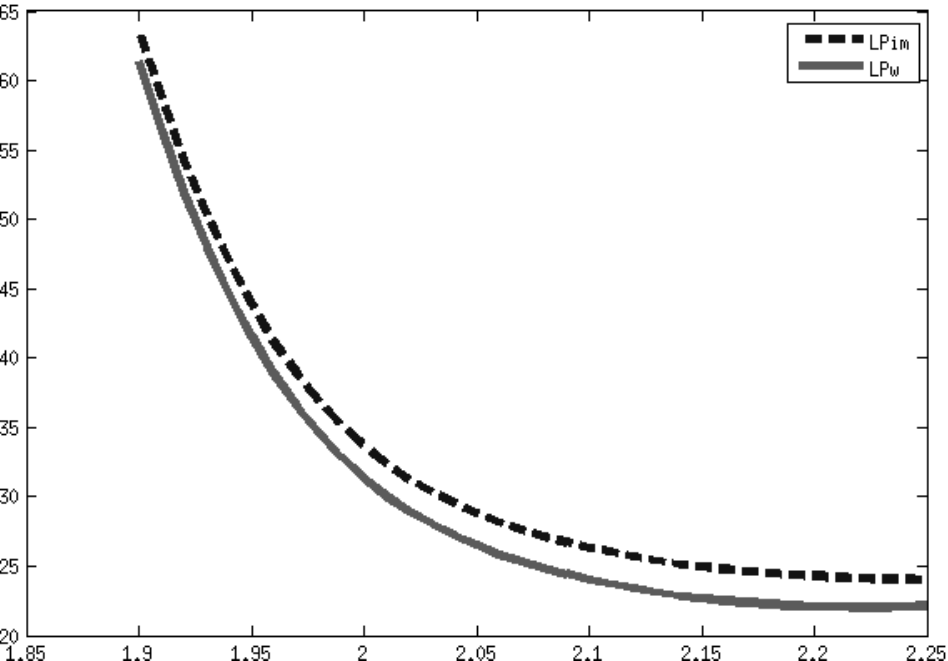}
\caption{Error curves $\|\Delta_j[w](x)-c(x)\cos(\omega_2x+\varphi_2)\|_{L^1}$ (solid line) and $\|\Delta_j[f](x)-c(x)\cos(\omega_2x+\varphi_2)\|_{L^1}$ 
(dashed line) with respect to $\omega_2$.}
\label{fig:l1curves}
\end{figure}

Then the component $w$ given by the decomposition algorithm and its Little\-wood-Paley filtered version $\Delta_j[w](x)$ are given in Fig.~\ref{fig:w}. 
To better understand what really happens, we can compare the Fourier transforms of $\Delta_j[w](x)$ and $\Delta_j[f](x)$. 
Figure~\ref{fig:ffts} shows the logarithm of the amplitude of these Fourier transforms. If we look carefully, we can see that some coefficients (mainly on the frequency axis)
 due notably to 
$a(x)$ are removed when we use $w$. This means that the extracted textures from $w$ are less affected by frequencies due to 
objects in the image (indeed, an object with sharp edges has high frequencies which add up to texture frequencies).\\
To completely verify the theorem, we sweep $\omega_2$ in a range which remains in the support covered by $\Delta_j$ and, in Fig.~\ref{fig:l1curves}, we 
plot the curves $\|\Delta_j[w](x)-c(x)\cos(\omega_2x+\varphi_2)\|_{L^1}$ (solid line) and $\|\Delta_j[f](x)-c(x)\cos(\omega_2x+\varphi_2)\|_{L^1}$ (dashed line).
Then we show that the decreasing error follows what is announced by the theorem (in $O(1/\omega_2)$). We can also see that the error in retrieving the
textures is lower if we filter $w$ instead of the original image directly. 

\section{Noisy images}\label{sec:noisy}
Now let us assume that $f(x)=a(x)+b(x)\cos(\omega x+\varphi)+\sigma R_N(x)$ where $R_N$ is a white noise filtered at a cutoff frequency $N$ (such that $f$ 
is of finite energy because the energy of a non-filtered white noise would be infinite). To simplify the problem we work on the bidimensional torus.

The $G-$norm of $R_N$ is $O(\sqrt{\log N})$. Then we suppose that $\sigma\sqrt{\log N}\simeq\lambda$ is $\lambda\leqslant\sqrt{\log N}$ 
(otherwise $\sigma=1$). We can use the same arguments as previously in the case where $\frac{\mu}{\lambda}\ll N, \lambda^2\ll \mu$. Then $f=u+v+w$ where 
$\|u\|_{BV}\leqslant C\lambda$ and $\|v\|_{BV}\leqslant C\frac{\mu}{\lambda}$ (because if we use this decomposition we have: 
$\|a\|_{BV}+\lambda\|b\|_2^2+\sigma\|R_N\|_G\leqslant C'\lambda$; consequently the optimal decomposition verifies 
$\|u\|_{BV}\leqslant \|u\|_{BV}+\lambda\|v\|_2^2+\mu\|w\|_G\leqslant C\lambda$; moreover we have $\|f\|_{BV}\geqslant\sigma N^2\sqrt{\log N}$ 
which corresponds to the second case of theorem \ref{theo2} where $\|v\|_{BV}\leqslant\frac{\mu}{2\lambda}$).

This gives
\begin{equation}
\|f-w\|_{BV}\leqslant C'\frac{\mu}{\lambda}
\end{equation}
and we can close the discussion like in section \ref{sec3}. We must use a bandpass filter which removes frequencies of the order of $|\omega|$ 
and keeps the essential energy of the filtered white noise. The Fourier series representation of a Gaussian white noise is given by 
$R(x)=\sum_k\sum_l g_{k,l}(\omega)e^{\imath (kx+ly)}$ where the coefficients $g_{k,l}(\omega)$ are independent identically distributed $N(0,1)$. 
Then we have,
\begin{equation}
R_N(x)=\sum_{|k|\leqslant N}\sum_{|l|\leqslant N}g_{k,l}e^{\imath(kx+ly)}
\end{equation}
and the Fourier coefficients of a function $g\in BV(\TD)$ verify $|\hat{g}(\boldsymbol{k})|\leqslant\frac{C}{|\boldsymbol{k}|}, \boldsymbol{k}=(k,l)$. 
There is a clear separation between the truncated white noise and the $BV$ function. This discussion can be resumed by the following lemma:

\begin{lemma}
Suppose we are given the sum $w(x)=R_N(x)+\eta(x)$ where $\|\eta\|_{BV}\leqslant\epsilon N$. Then $w(x)$ remains close 
to a white noise of cutoff frequency $N$ in the sense that, if $|k|\leqslant N, |l|\leqslant N, \boldsymbol{k}=(k,l)$, we have 
$|\hat{w}(\boldsymbol{k})-g_{k,l}(\omega)|\leqslant\epsilon\frac{N}{|\boldsymbol{k}|}$.
\end{lemma}

\section{Multiscale texture separation (MTS) algorithm}\label{sec:textsep}
\subsection{Algorithm description}
\begin{figure}[!t]
\includegraphics[width=\textwidth]{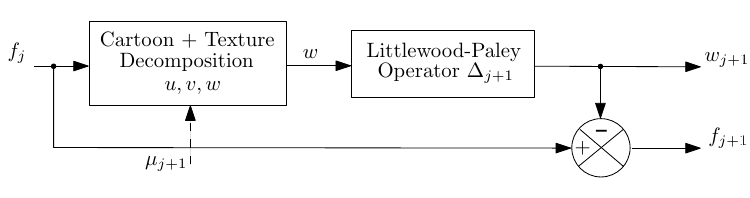}
\caption{Single scale texture separation block}
\label{fig:sstsb} 
\end{figure}

\begin{figure}[!t]
\includegraphics[width=\textwidth]{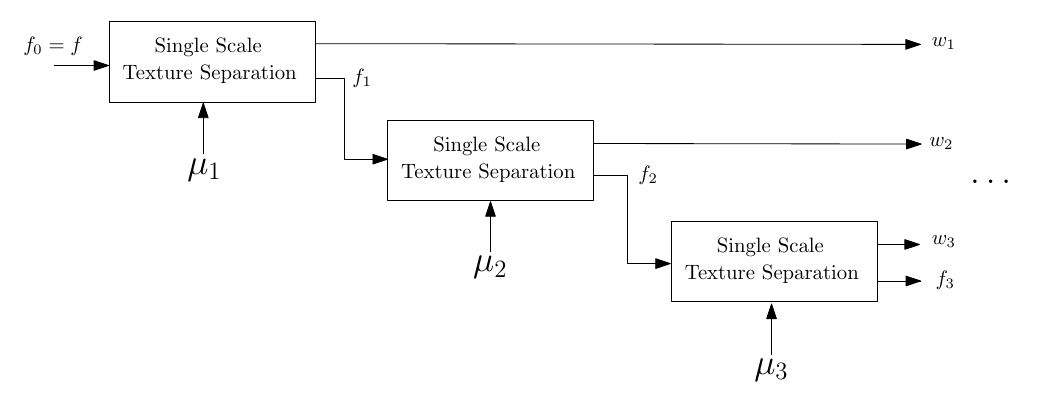}
\caption{Multiscale scale texture separation}
\label{fig:msts}
\end{figure}

If Theorem.~\ref{th:sep} is useful to extract specific textures, it assumes that we must know in which shell lies the texture we want to retrieve. Indeed this knowledge is necessary
to properly fix the parameters $\mu,\lambda$ and the scale $j$. However, for texture analysis purposes, such information is generally unknown. Then, we propose 
to design a multiscale texture separation algorithm by extracting recursively the textures corresponding to different scales. \\
In such use of the decomposition, Theorem.~\ref{th:sep} tells us that we can fix $\lambda$ to one, and the idea is to start by choosing the scale $j$ which 
corresponds to the highest frequencies. Then we can use the Theorem to fix the parameter $\mu$ associated with this scale. We compute the cartoon + 
texture decomposition and finally apply the Littlewood-Paley operator to accurately extract the most oscillating textures. Let us denote $w_j=\Delta_j[w]$. 
The lower oscillating counterpart is obtained by substracting $w_{j+1}$ from the input image $f_j$. In the sequel we denote $f_{j+1}=f_j-w_{j+1}$, then
a single scale texture separation block can be depicted as in 
Fig.~\ref{fig:sstsb}. This process can be iterated to reach a multiscale texture separation algorithm (because we consider dyadic scales, we can set $\mu_{j+1}=\mu_j/2$) as in
Fig.~\ref{fig:msts}. The following lemma is an obvious property of this multiscale decomposition.
\begin{proposition}
Let $J$ denote the number of scales, $\{f_j,w_j\}_{j\in[1,J]}$ the set of components obtained by the multiscale texture separation of the original image $f$, 
then
\begin{equation}
 f=f_J+\sum_{j=1}^Jw_j
\end{equation} 
\end{proposition}
\begin{proof}
It is straightforward from the construction of the multiscale texture separation algorithm.
\qquad\end{proof}

\subsection{Examples}
\begin{figure}[!t]
\centering $f_0=f$\\
\centering\includegraphics[width=0.35\textwidth]{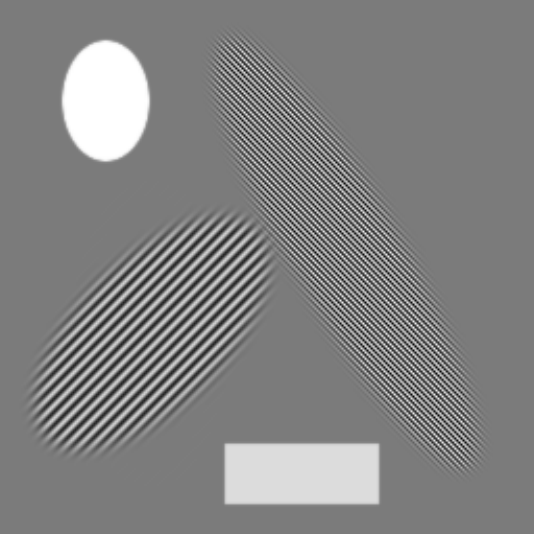}\\[1mm]
\begin{tabular}{cc}
\includegraphics[width=0.35\textwidth]{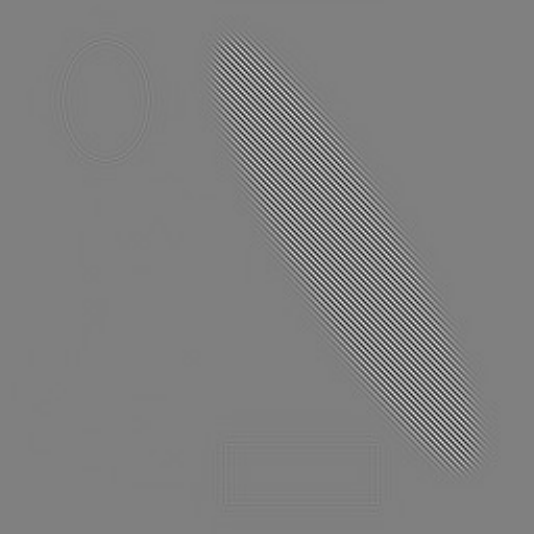} & \includegraphics[width=0.35\textwidth]{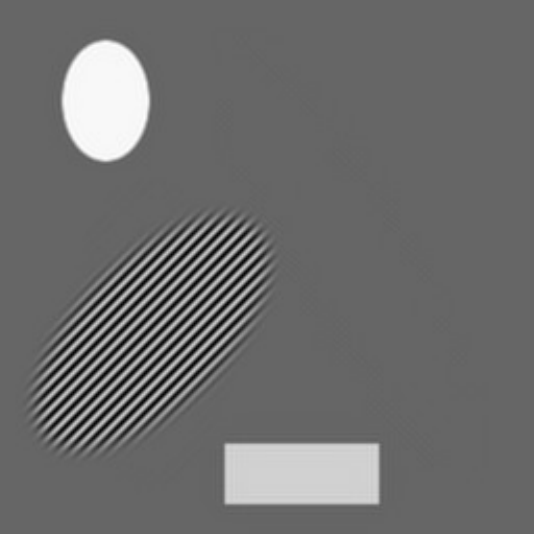} \\
$w_1$ & $f_1$ \\
\includegraphics[width=0.35\textwidth]{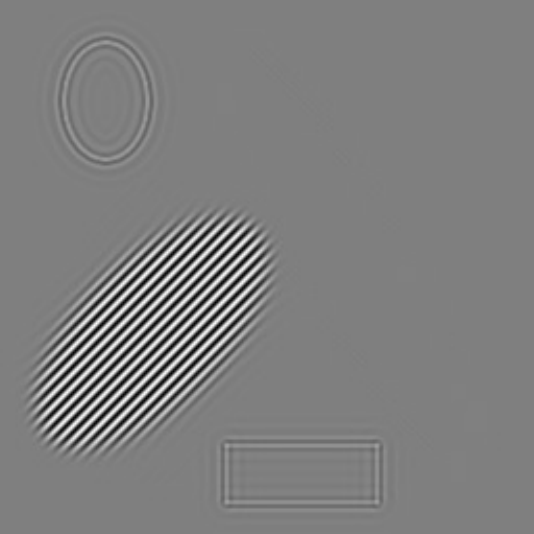} & \includegraphics[width=0.35\textwidth]{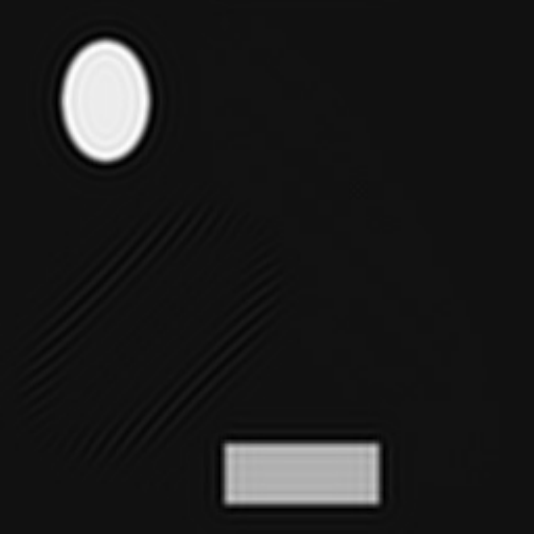} \\
$w_2$ & $f_2$ \\
\includegraphics[width=0.35\textwidth]{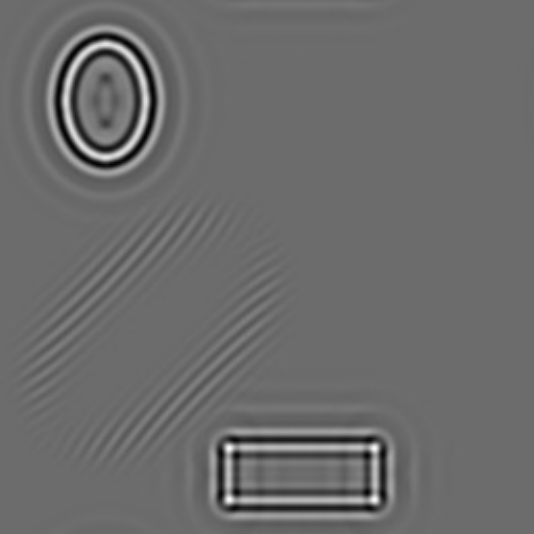} & \includegraphics[width=0.35\textwidth]{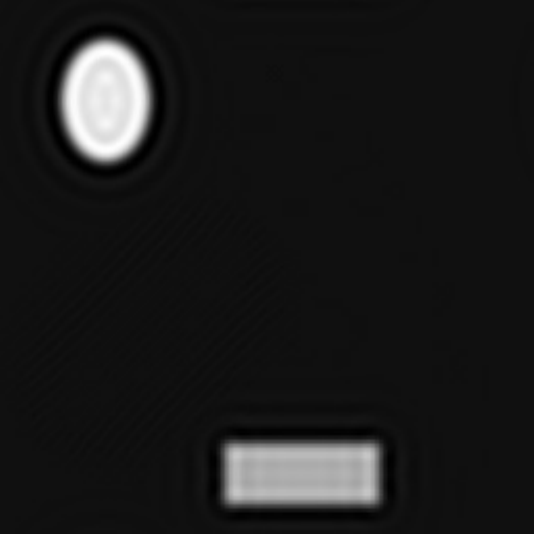} \\
$w_3$ & $f_3$
\end{tabular} 
\caption{Example of the first three scales components obtained from a synthetic image.}
\label{fig:exmts}
\end{figure}

In Fig.~\ref{fig:exmts}, we present the result of the multiscale texture separation applied on a synthetic image. The input image is created in such a way that 
the two oscillating components have their frequencies located in two consecutive shells, respectively. On the obtained components, $w_1$ and $w_2$, we can see that the different texture 
parts are well separated. Some ``ringing'' artifacts can be observed (see for example components $w_2$ and $f_2$ in Fig.~\ref{fig:exmts}). 
These effects are coming from two facts. Firstly, the decomposition doesn't perfectly extract sharp objects and some Fourier coefficients, mainly corresponding to edges, are still in the Fourier spectrum 
of the texture part and then captured by the Littlewood-Paley filter. These phenomena can be observed in Fig.~\ref{fig:ffts} where some coefficients, coming from the cartoon part, remain on the 
vertical and horizontal axis. Secondly, there is no guarantee that a texture has its Fourier support belonging to a single scale. Indeed, the size of its Fourier support depends directly on the regularity of the functions 
$b(x)$ and $c(x)$ and can lie in contiguous scales. 

\begin{figure}[!t]
\centering
\begin{tabular}{cc}
\includegraphics[width=0.48\textwidth]{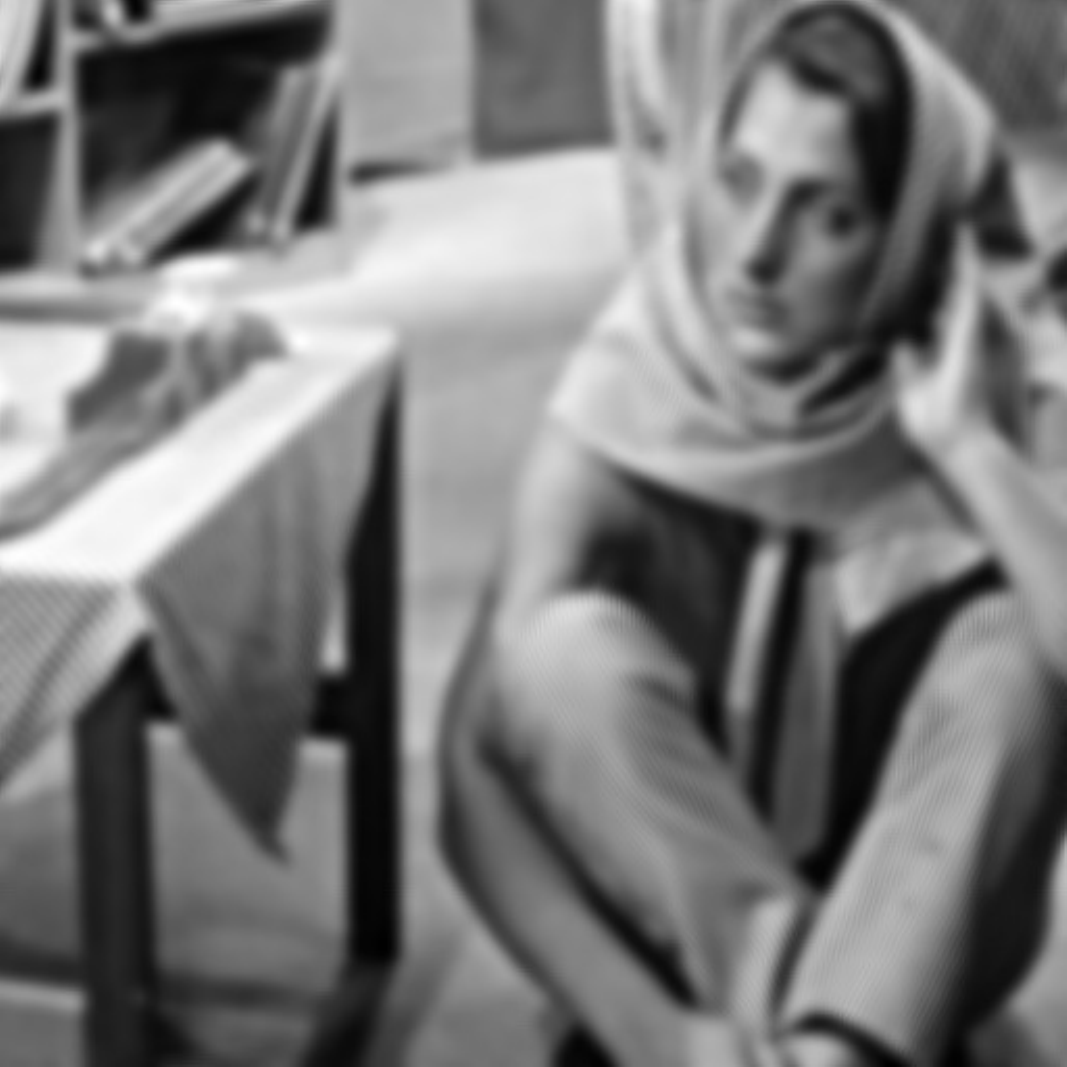} & \includegraphics[width=0.48\textwidth]{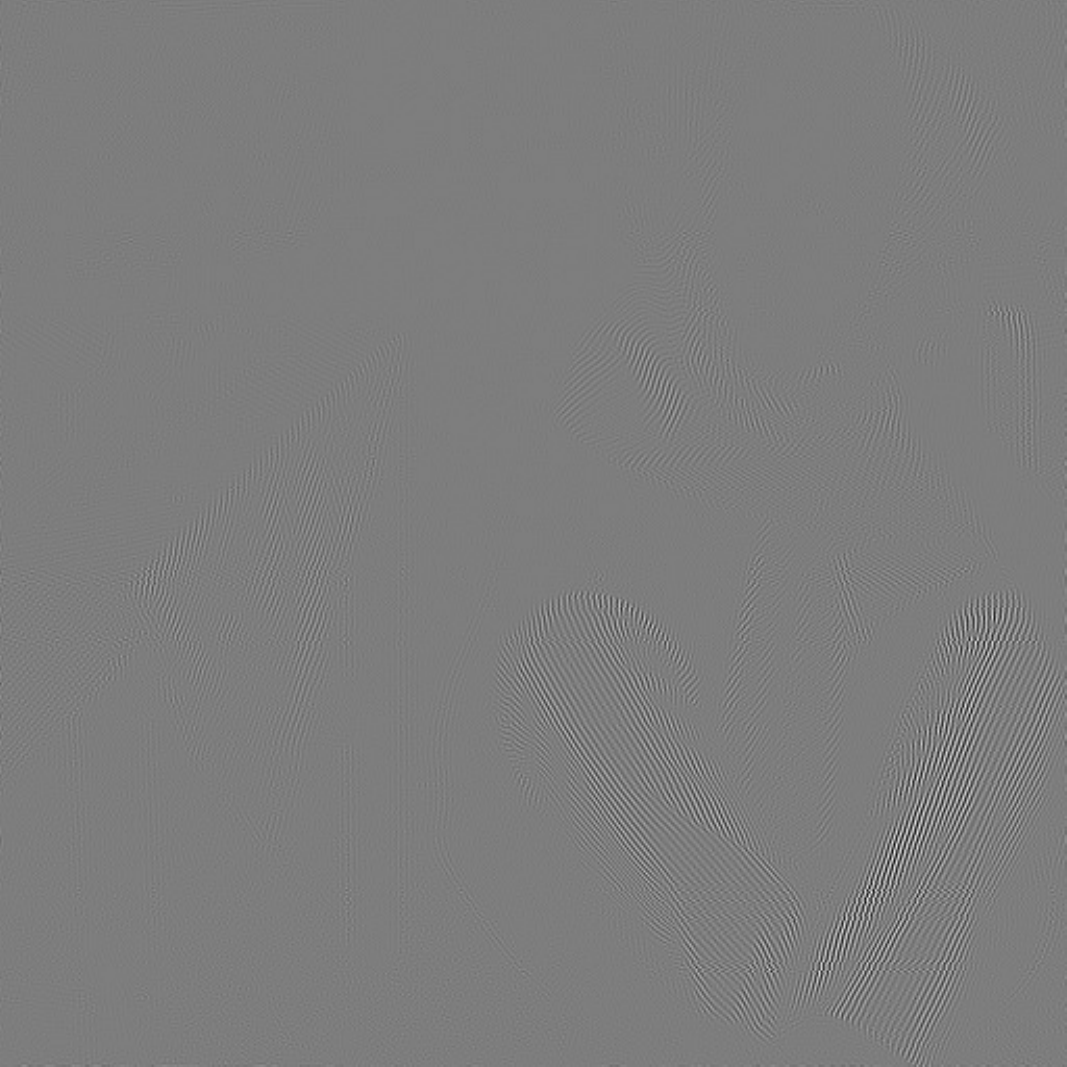} \\
$f_4$ & $w_1$ \\
\includegraphics[width=0.48\textwidth]{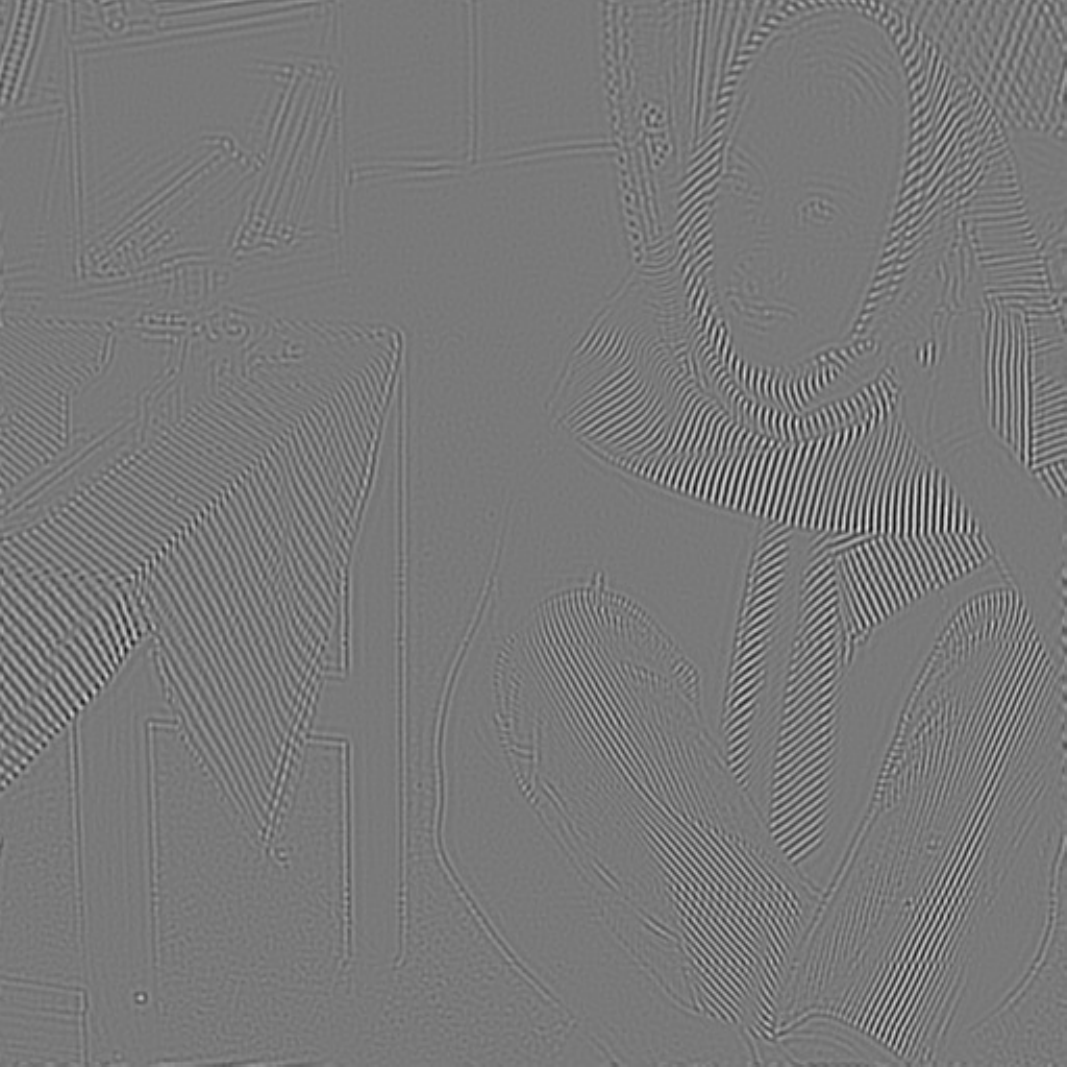} & \includegraphics[width=0.48\textwidth]{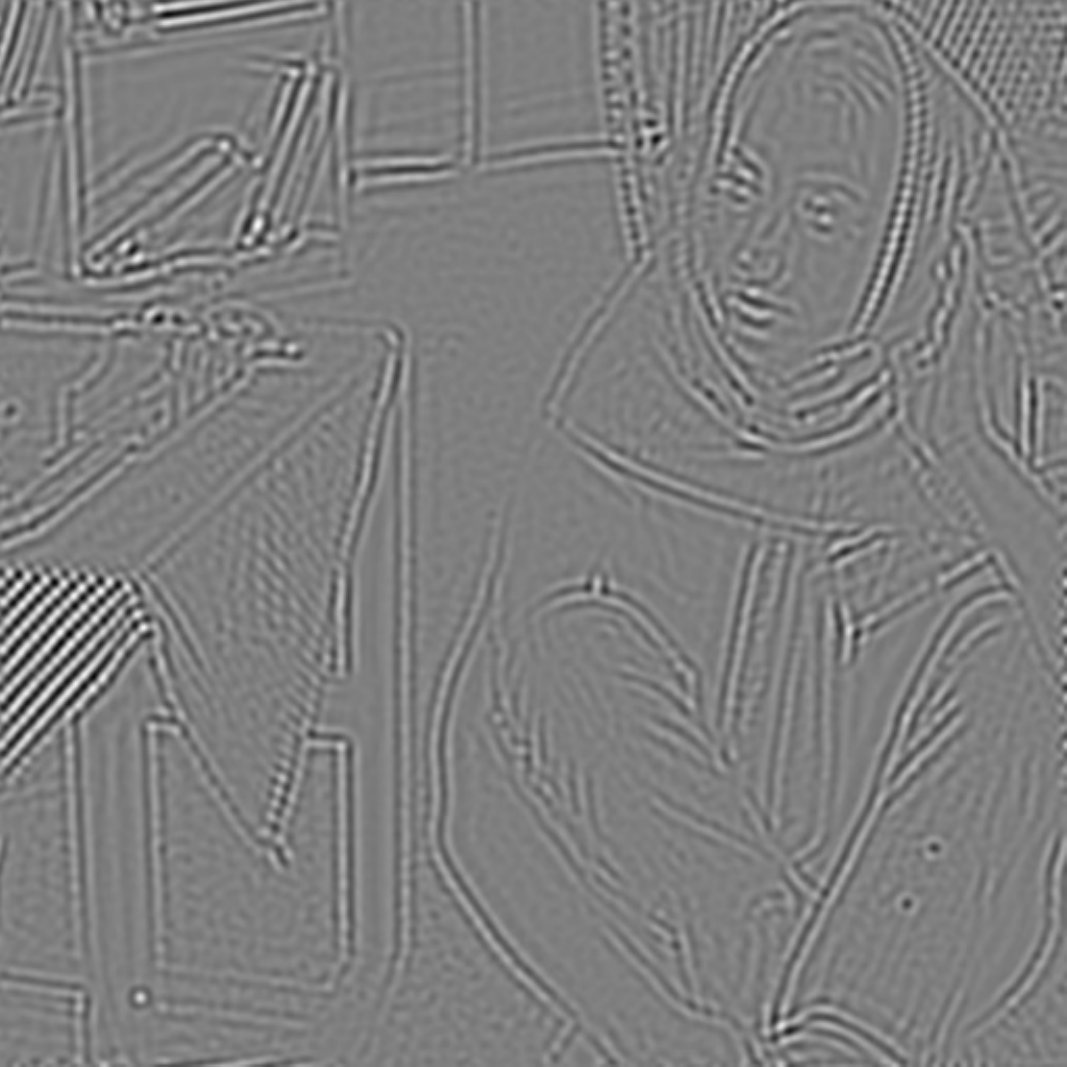} \\
$w_2$ & $w_3$ \\
\includegraphics[width=0.48\textwidth]{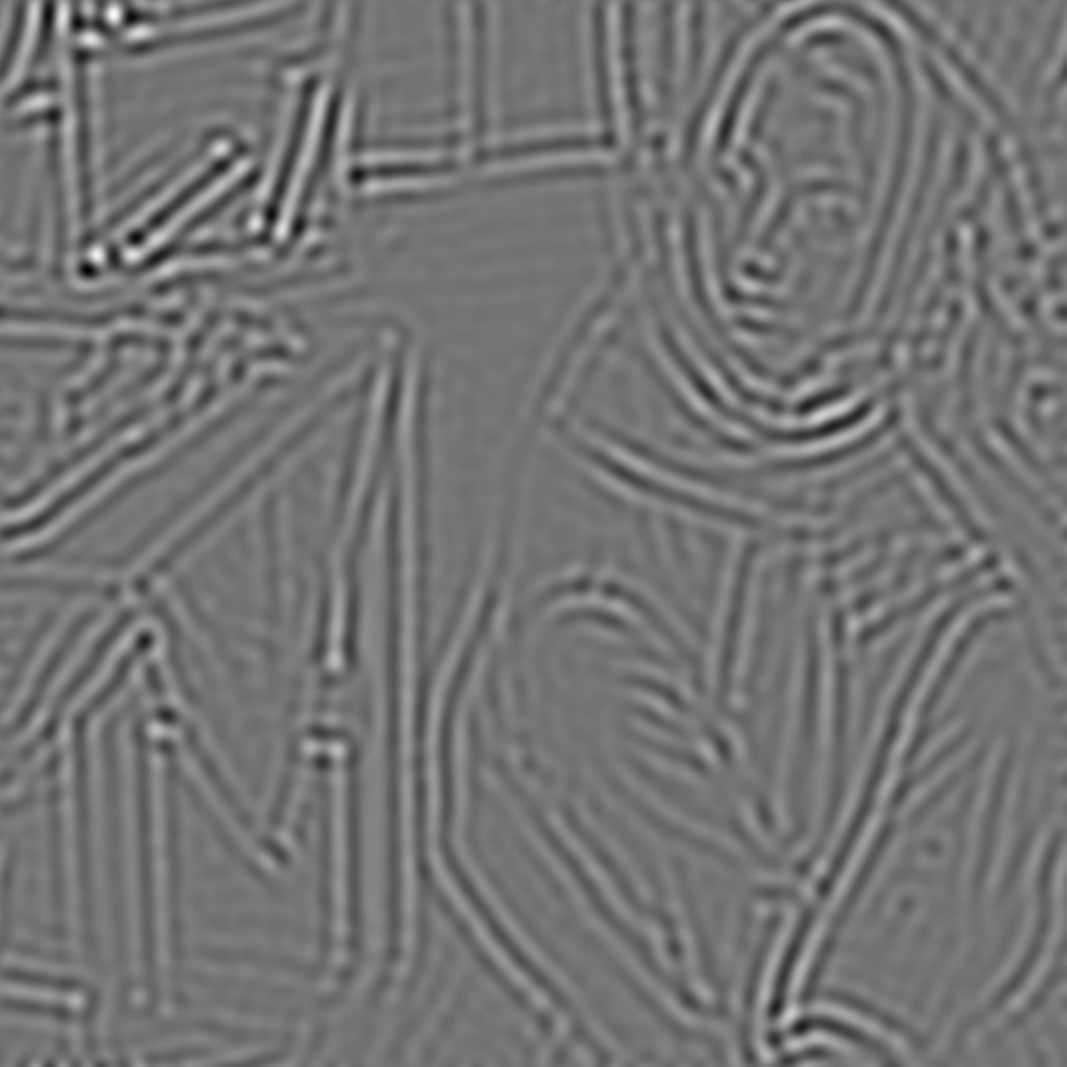} & \includegraphics[width=0.48\textwidth]{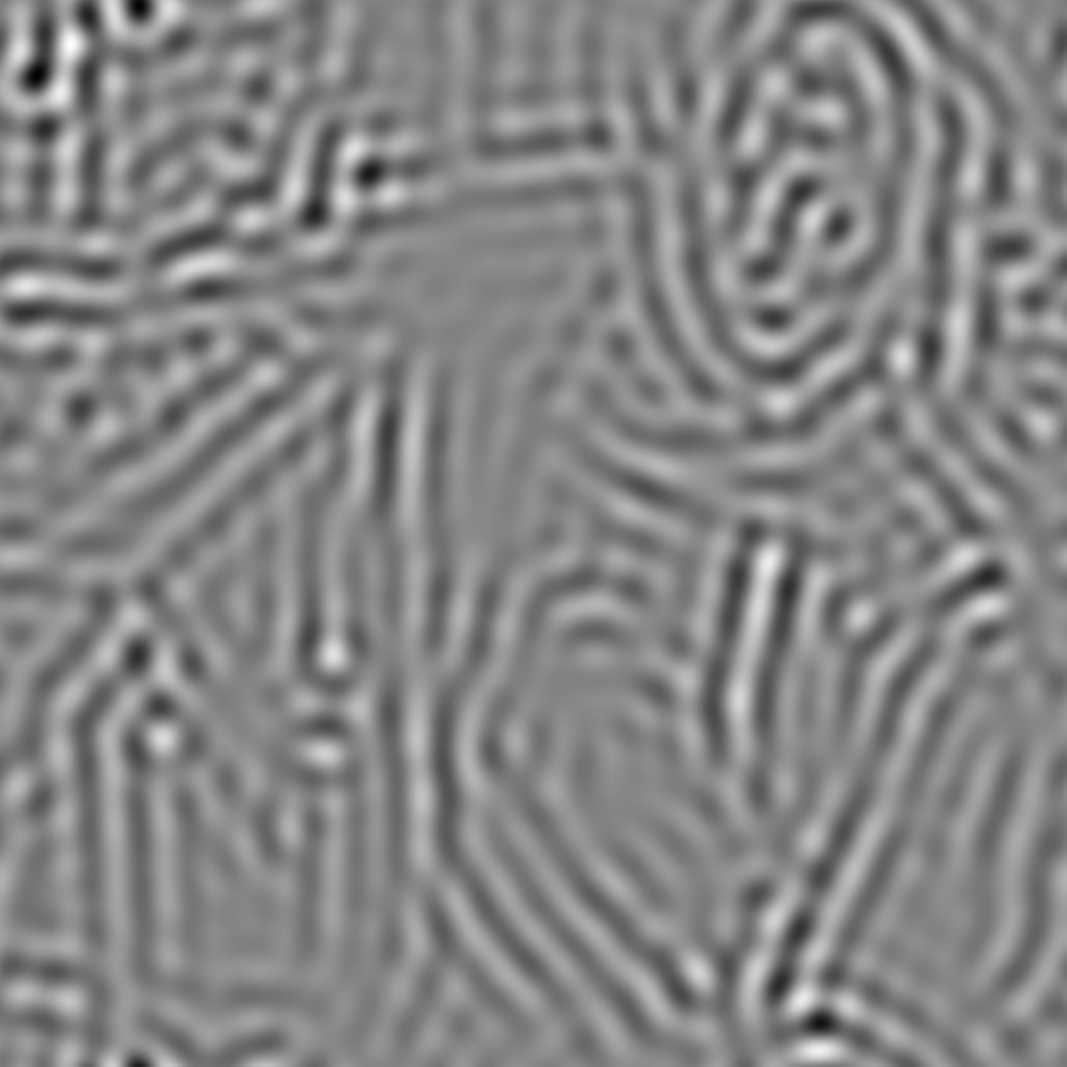} \\
$w_4$ & $w_5$
\end{tabular}
\caption{Five scales decomposition of Barbara image.}
\label{fig:barbmts}
\end{figure}

A multiscale texture separation on a real image is given in Fig.~\ref{fig:barbmts}. We can observe that the algorithm captures well the textures corresponding to different scales, $w_1$ has the most oscillating textures, 
$w_2$ less oscillating ones and so on in the consecutive components. This arises the question ``how many levels of decomposition must we choose?''. In some sense, this question is the same as for the wavelet transform, we 
a priori do not know the best expansion depth. The coefficient $J$ remains a parameter of the algorithm. It is natural to think that the choice of this parameter will depend on the kind of image we want to analyze.
Further investigation is needed to find a way to estimate the best $J$.

\section{Directional MTS}\label{sec:dmts}
\begin{figure}[!t]
\includegraphics[width=\textwidth]{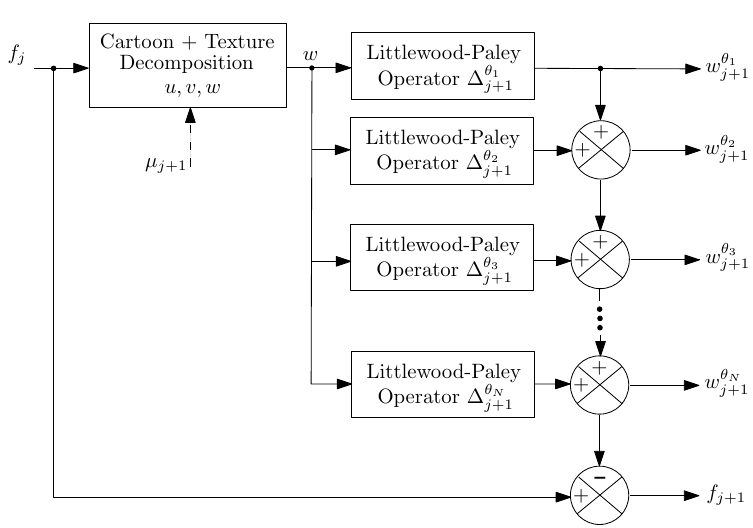}
\caption{Directional Single scale texture separation block}
\label{fig:sstsbd} 
\end{figure}

Usually, in texture analysis, one uses the orientation information to build discriminating features. In the case of our multiscale texture separation 
algorithm, an easy way to add such functionality is to consider only portions of each dyadic shell. This idea was previously used in the construction of curvelet 
frames by Cand\`es et al. \cite{Candes2003a,Candes2003b,Candes2005}. Then we can build a directional filter bank following the construction of the curvelets 
by normalizing them in order to have an amplitude of one in their domain of definition to match the properties of Littlewood-Paley filters. Now, instead of applying 
only one filter to the texture part, we apply this filter bank and get the different textures corresponding to each direction. This corresponds to modifying 
the initial single texture separation block shown in Fig.~\ref{fig:sstsb} into a directional single texture separation block as depicted in Fig.\ref{fig:sstsbd} 
where $\Delta_j^{\theta_l}$ represents the Littlewood-Paley filter at scale $j$ associated with direction $\theta_l$.

\begin{figure}[!t]
\centering
\begin{tabular}{cc}
\includegraphics[width=0.47\textwidth]{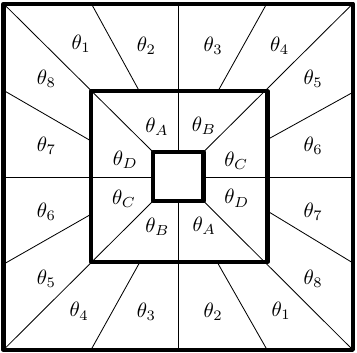} & \includegraphics[width=0.47\textwidth]{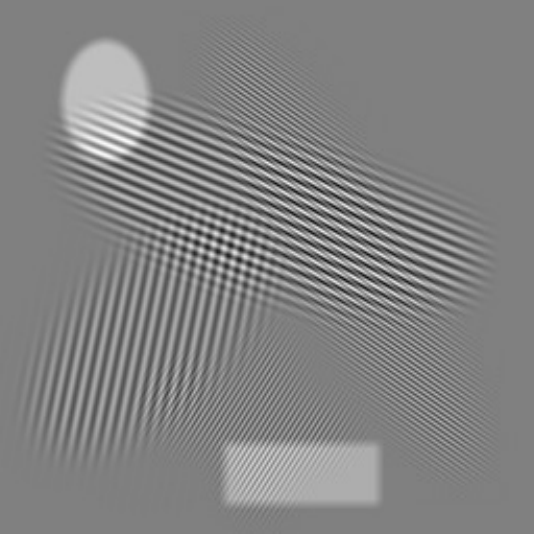} \\
(a) & (b)
\end{tabular}
\includegraphics[width=0.47\textwidth]{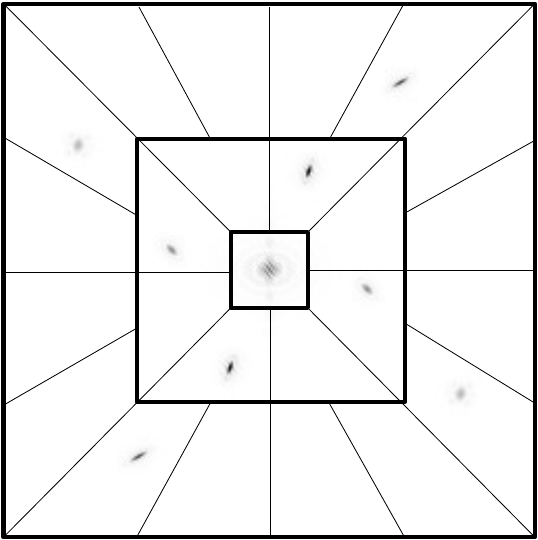}\\
(c)
\caption{Partition of the Fourier domain (a), test image (b) and its Fourier transform (c).}
\label{fig:syndmts}
\end{figure}

\begin{figure}[!t]
\centering\begin{tabular}{cc}
\includegraphics[width=0.35\textwidth]{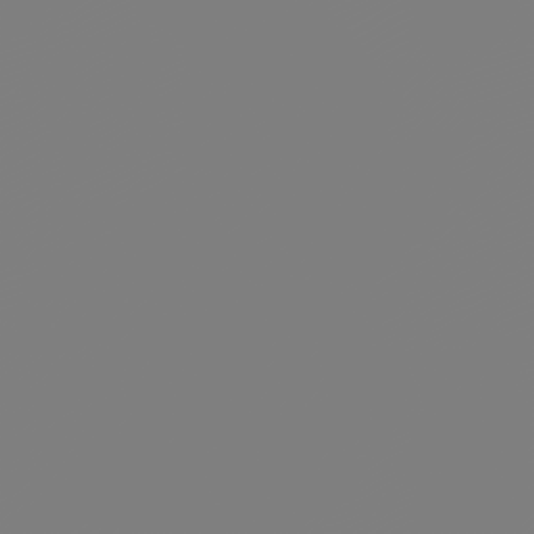} & \includegraphics[width=0.35\textwidth]{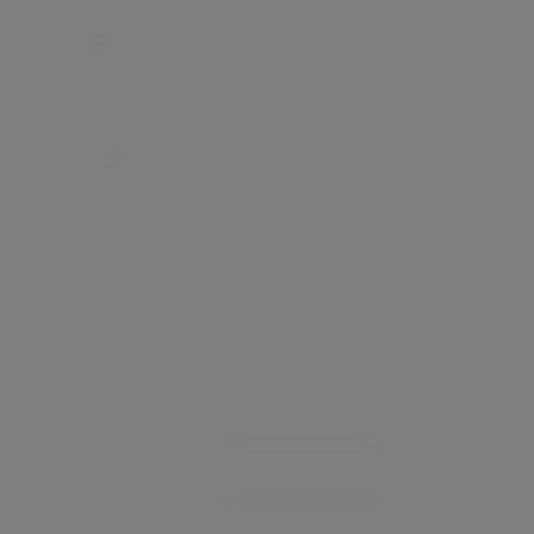} \\
$\theta_1$ & $\theta_2$ \\
\includegraphics[width=0.35\textwidth]{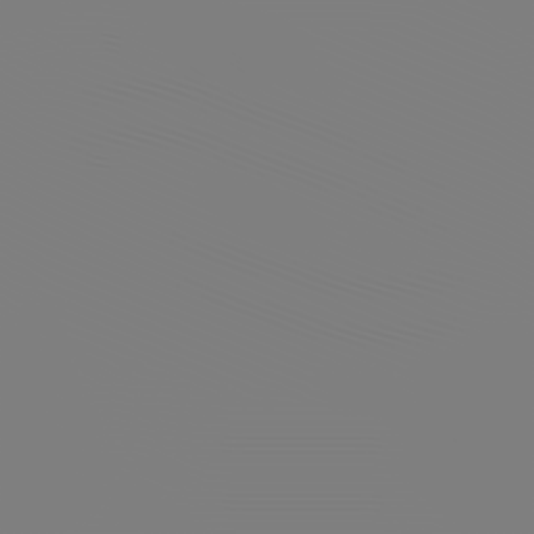} & \includegraphics[width=0.35\textwidth]{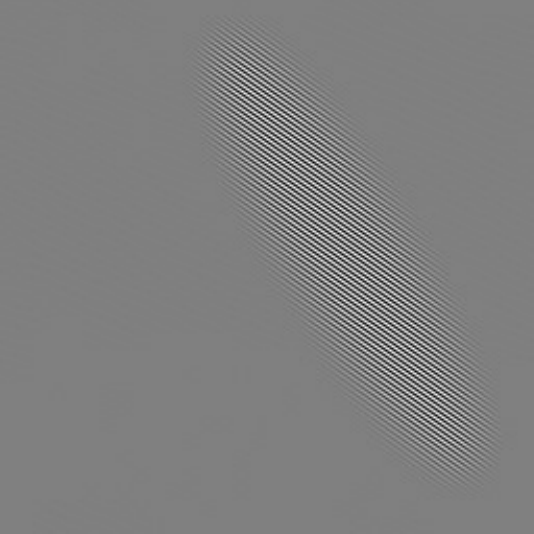} \\
$\theta_3$ & $\theta_4$ \\
\includegraphics[width=0.35\textwidth]{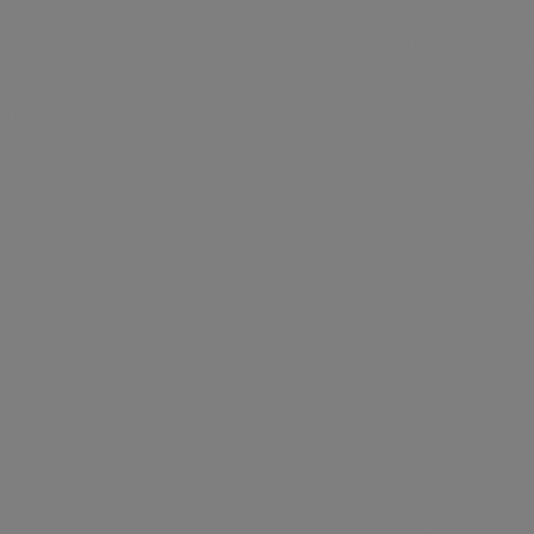} & \includegraphics[width=0.35\textwidth]{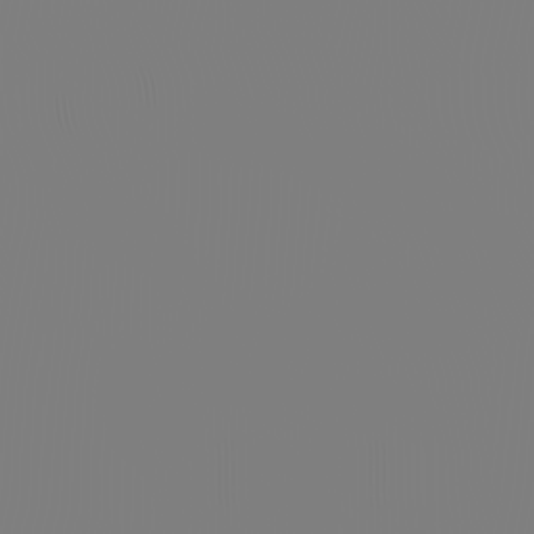} \\
$\theta_5$ & $\theta_6$ \\
\includegraphics[width=0.35\textwidth]{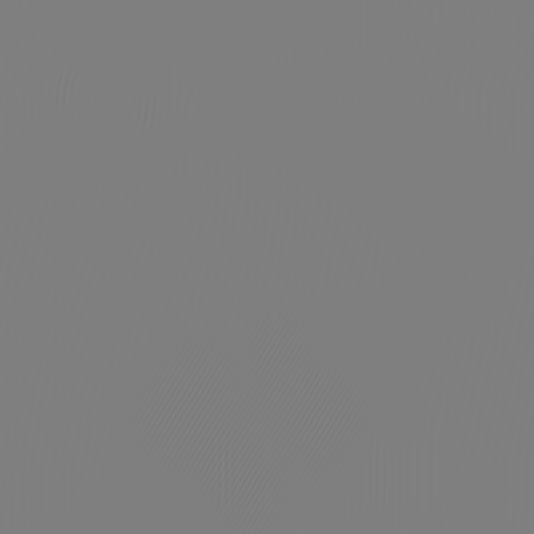} & \includegraphics[width=0.35\textwidth]{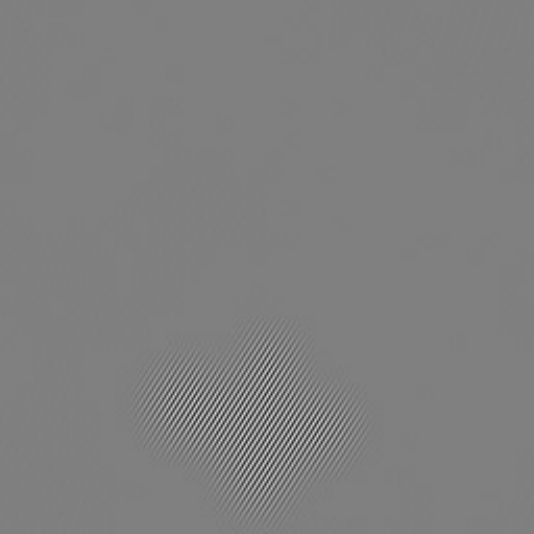} \\
$\theta_7$ & $\theta_8$
\end{tabular} 
\caption{First scale directional texture components obtained from the synthetic image.}
\label{fig:S1dmts}
\end{figure}

\begin{figure}[!t]
\centering\begin{tabular}{cc}
\includegraphics[width=0.36\textwidth]{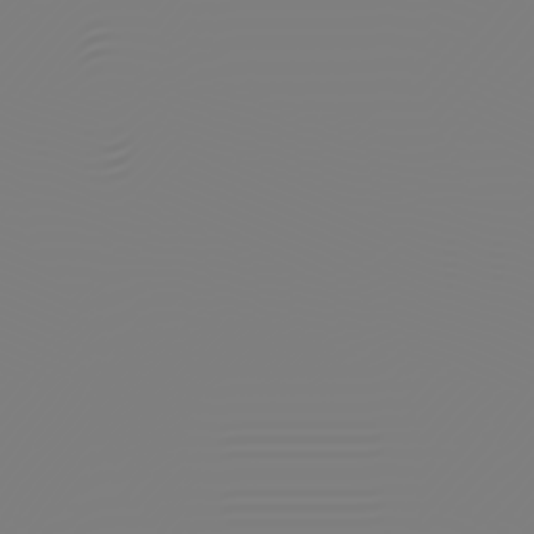} & \includegraphics[width=0.36\textwidth]{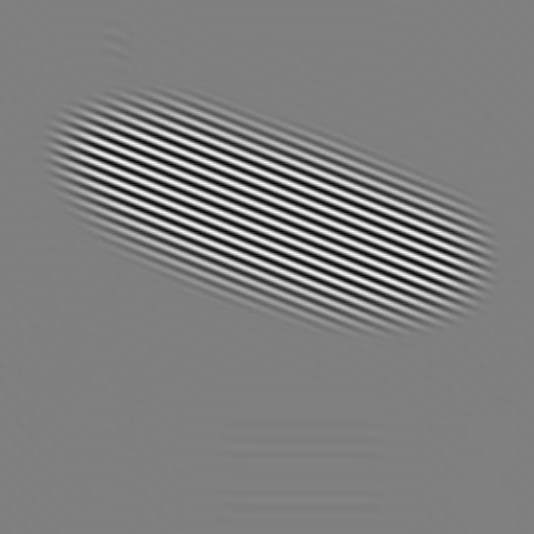} \\
$\theta_A$ & $\theta_B$ \\
\includegraphics[width=0.36\textwidth]{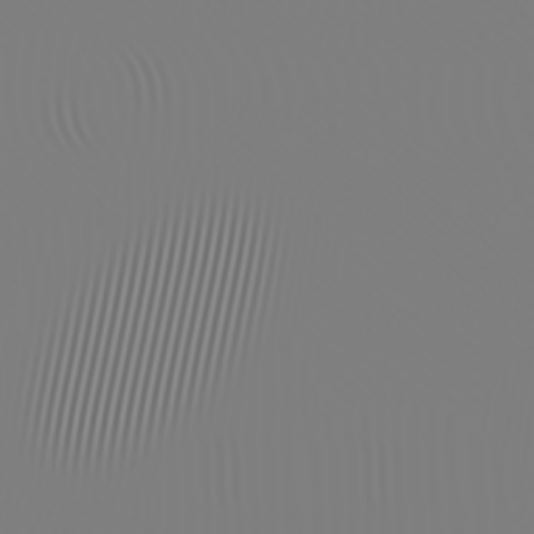} & \includegraphics[width=0.36\textwidth]{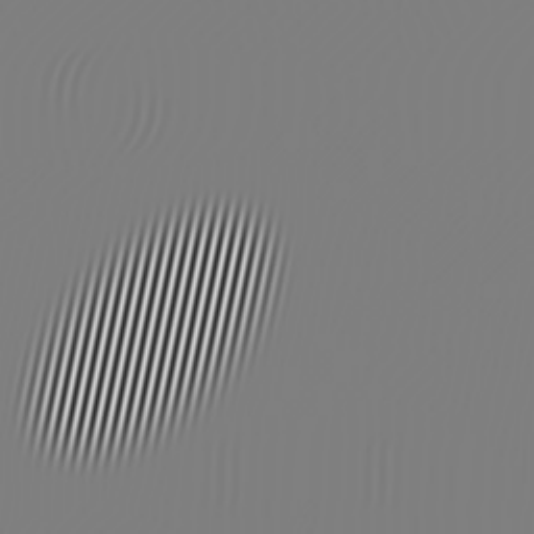} \\
$\theta_C$ & $\theta_D$ \\
\end{tabular}
\includegraphics[width=0.36\textwidth]{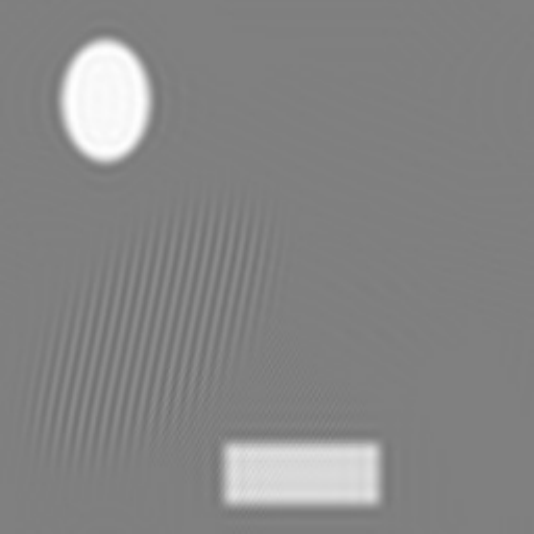} \\
$f_2$
\caption{Second scale directional texture components and the low scale image $f_2$ obtained from the synthetic image.}
\label{fig:S2dmts}
\end{figure}

We test this directional MTS algorithm by setting a partition of the Fourier domain like the one depicted in Fig.~\ref{fig:syndmts}.a. We fix two scales 
(delimited by the bold squares) and eight directions for the outer shell and four directions for the inner shell. Then we build a test image 
(Fig.\ref{fig:syndmts}.b) composed of a cartoon part and four different textures, two of them lie in the outer shell and the other two in the inner shell, 
all with different orientations. Fig.\ref{fig:syndmts}.c shows its Fourier transform and the localization (with respect to the partitioning) of the main 
coefficients corresponding to each component. Figures~\ref{fig:S1dmts} and \ref{fig:S2dmts} give the output of the directional MTS algorithm. We can see that, 
as expected from the Fourier transform of the input image, textures focused in directions $\theta_4, \theta_8, \theta_B$ and $\theta_D$ are well 
separated on their corresponding components.

\section{Conclusion}\label{sec:conc}
In this paper we proved that a Littlewood-Paley filtering of the texture part, coming from a $BV-G$ image decomposition, permits to 
almost perfectly extracting oscillating components of an image. Based on this result we built a multiscale texture separation algorithm. This 
algorithm permits to extract textures which oscillate at different scales. Finally an extension from this multiscale texture separation algorithm into a 
directional MTS was proposed. This version allows us to separate textures which are close in terms of frequencies but have different orientations.\\
The different experiments show that the outputs of the algorithms follow the predicted results of the theory and could be useful for texture analysis 
purposes.\\
Further investigation is currently underway in order to generalize this approach to build an adaptive decomposition algorithm.

\section*{Acknowledgments}
The author wants to thank Prof. Yves Meyer for his useful advice and his involvement in this work and Prof. Stanley Osher for his support. The author also wants to 
thank the referee for their comments and suggestions which permit to improving the manuscript quality.


\clearpage
\begin{thebibliography}{10} 
\bibitem{Aujol2005a} {\sc Jean-Fran\c{c}ois Aujol and Gilles Aubert and Laure Blanc-F\'eraud and Antonin Chambolle},
  {\em Image decomposition into a bounded variation component and an oscillating component},
  Journal of Mathematical Imaging and Vision, Vol.~22, No.~1, pp.~71--88, 2005.

\bibitem{Aujol2003a} {\sc Jean-Fran\c{c}ois Aujol and Gilles Aubert and Laure Blanc-F\'eraud and Antonin Chambolle},
  {\em Decomposing an image. Application to textured images and {SAR} images},
  Technical Report, Universit\'e de Nice Sophia-Antipolis, 2003.

\bibitem{Aujol2005b} {\sc Jean-Fran\c{c}ois Aujol and Antonin Chambolle},
  {\em Dual norms and image decomposition models},
  International Journal of Computer Vision, Vol.~63, No.~1, pp.~85--104,2005.

\bibitem{Aujol2006a} {\sc Jean-Fran\c{c}ois Aujol and Tony Chan},
  {\em Combining geometrical and textured information to perform image classification},
  Journal of Visual Communication and Image Representation, Vol.~7, pp.~1004--1023, 2006.

\bibitem{Aujol2006b} {\sc Jean-Fran\c{c}ois Aujol and Guy Gilboa and Tony Chan and Stanley Osher},
  {\em Structure-texture image decomposition-modeling, algorithms and parameter selection},
  International Journal of Computer Vision, Vol.~67, No.~1, pp.~111--136, 2006.

\bibitem{Aujol2006c} {\sc Jean-Fran\c{c}ois Aujol and S.H. Kang},
  {\em Color image decomposition and restoration},
  Journal of Visual Communication and Image Representation, Vol.~17, No.~4, pp.~916--928, 2006.

\bibitem{Aujol2004} {\sc Jean-Fran\c{c}ois Aujol and Basarab Matei},
  {\em Structure and texture compression},
  Technical Report, {INRIA}, No.~5076, 2004.

\bibitem{Aujol2004a} {\sc Jean-Fran\c{c}ois Aujol and Batasarab Matei},
  {\em Simultaneous structure and texture compact representation},
  Advanced Concepts for Imaging Vision Systems (ACIVS) Conference, 2004.

\bibitem{Gilles2007} {\sc J\'er\^ome Gilles},
  {\em Noisy image decomposition: a new structure, texture and noise model based on local adaptivity},
  Journal of Mathematical Imaging and Vision (JMIV), Vol.~28, No.~3, pp.~285--295, 2007.

\bibitem{Gilles2010a} {\sc J\'er\^ome Gilles and Yves Meyer},
  {\em Properties of BV-G structures + textures decomposition models. Application to road detection in satellite images},
  IEEE Transaction in Image Processing, Vol.~19, No.~11, pp.~2793--2800, 2010.

\bibitem{Gilles2012b} {\sc J\'er\^ome Gilles and Stanley Osher},
  {\em Bregman implementation of Meyer's G-norm for cartoon + textures decomposition},
  UCLA CAM Report 11-73, 2011.

\bibitem{Kurtz1987a} {\sc Douglas S. Kurtz},
  {\em Littlewood-Paley operators on BMO},
  Proceedings of the American Mathematical Society, Vol.~99, No.~4, pp.~657--666, 1987.

\bibitem{Triet2005a} {\sc Triet M. Le and Luminita A. Vese},
  {\em Image Decomposition Using Total Variation and div(BMO)},
  Multiscale Modeling and Simulation: A SIAM Interdisciplinary Journal, Vol.~4, No.~2, pp.~390--423, 2005.

\bibitem{Meyer2001} {\sc Yves Meyer},
  {\em Oscillating patterns in image processing and in some nonlinear evolution equations},
  The Fifteenth Dean Jacquelines B. Lewis Memorial Lectures, 2001.

\bibitem{Osher2003a} {\sc Stanley Osher and Andr\`es Sole and Luminita Vese},
  {\em Image decomposition and restoration using total variation minimization and the {H}$^{-1}$ norm},
  Multiscale Modeling and Simulation: A SIAM Interdisciplinary Journal, Vol.~1, No.~3, pp.~349--370, 2003.

\bibitem{Luminita2003a} {\sc Luminita Vese and Stanley Osher},
  {\em Modeling textures with total variation minimization and oscillating patterns in image processing},
  Journal of Scientific Computing, Vol.~19, pp.~553--572, 2003.

\bibitem{Rudin1992} {\sc Leonid Rudin and Stanley Osher and Emad Fatemi},
  {\em Nonlinear total variation based noise removal algorithms},
  Physica D, Vol.~60, pp.~259--268, 1992.

\bibitem{Candes2003a} {\sc Emmanuel Cand\`es and David Donoho},
  {\em Continuous Curvelet Transform, part {I}: Resolution of the Wavefront Set},
  Applied Computational Harmonic Analysis, Vol.~19, pp.~162--197, 2003.

\bibitem{Candes2003b} {\sc Emmanuel Cand\`es and David Donoho},
  {\em Continuous Curvelet Transform, part {II}: Discretization and Frames},
  Applied Computational Harmonic Analysis, Vol.~19, pp.~198--222, 2003.

\bibitem{Candes2005} {\sc Emmanuel Cand\`es and Laurent Demanet and David Donoho and Lexing Ying},
  {\em Fast discrete curvelet transforms},
  Multiscale Modeling and Simulation: A SIAM Interdisciplinary Journal, Vol.~5, pp.~861--899, 2005.

\bibitem{Tadmor2004} {\sc Eitan Tadmor and Suzanne Nezzar and Luminita Vese},
  {\em A Multiscale Image Representation Using Hierarchical $(BV,L^2)$ Decompositions},
  Multiscale Modeling and Simulation: A SIAM Interdisciplinary Journal, Vol.~2, No~4, pp.~554--579, 2004.
 
\bibitem{aubin1984applied} {\sc Jean-Pierre Aubin and Ivar Ekeland},
  {\em Applied nonlinear analysis},
  Pure and applied mathematics: Wiley-Interscience series of texts, monographs and tracts, ISBN-9780471059981, 1984.

\end{thebibliography}
\end{document}